\documentstyle[prl,twocolumn,aps,epsfig]{revtex}

\begin{document}
\draft
\twocolumn[\hsize\textwidth\columnwidth\hsize\csname @twocolumnfalse\endcsname
\title{Terrace-Width Distributions and Step-Step Repulsions on Vicinal Surfaces:
Symmetries, Scaling, Simplifications, Subtleties, and Schr\"odinger
}
\vspace{-0.3cm}
\author{T.~L.\ Einstein$^*$,
Howard L.\ Richards$^{\dag}$, Saul~D.\ Cohen, and O.
Pierre-Louis$^{\ddag}$}      

\address{Department of Physics,
University of Maryland, College Park, MD 20742-4111 USA\\}

\date{\today}
\maketitle
\begin{abstract}
For more than three decades, measurement of  terrace width distributions (TWDs) of
vicinal crystal surfaces have been recognized as arguably the best way to determine the
dimensionless strength $\tilde{A}$  of the elastic repulsion between steps. For
sufficiently strong repulsions, the TWD is expected to be Gaussian, with
$\tilde{A}$ varying inversely with the squared variance.  However, there has been a
controversy over the proportionality constant.  From another perspective the TWD can
be described as a continuous generalized Wigner distribution (CGWD) essentially no more
complicated than a Gaussian but a much better approximation at the few calibration
points where exact solutions exist. This paper combines concisely the experimentally
most useful results from several earlier papers on this subject and describes some
advancements that are in progress regarding numerical tests and in using
Schr\"odinger-equation formalism to give greater understanding of the origin of the CGWD
and to give hope of extensions to more general interaction potentials between
steps.  There are many implications for future experiments.
\end{abstract}
]

\section{Introduction}
\label{sec:intro}

Quantitative measurement of the widths $\ell$ of terraces on vicinal surfaces
became possible a decade ago. A principal motivation for examining the terrace
width distribution (TWD) is the recognition that it provides arguably the
optimal way to assess the strength
 of the elastic (and/or dipolar) repulsion between steps, specifically the
coefficient
$A$ of the elastic repulsion per length $A/\ell^2$.  Here the elastic
repulsion is taken to be  perpendicular to the mean step direction.
 All standard analysis procedures make a continuum approximation in
the direction along the steps, called $\hat{y}$ in ``Maryland
notation." (The perpendicular direction in the terrace plane, in the
``downstairs" direction, is denoted $\hat{x}$.)  Hereafter,
$A$ appears only in form of a {\em dimensionless interaction strength}
\begin{equation}
  \label{e:defAtilde}
    \tilde{A} \equiv A\tilde{\beta}(k_BT)^{-2}\; ,
\end{equation}
where $\tilde{\beta}$ is the step stiffness.  

Experimentally, a TWD is typically characterized by
its variance $\sigma^2$ and, at least when $\tilde{A}$ is not small,
has a shape that can be satisfactorily approximated by a Gaussian.  The
Gaussian form can be readily derived from a mean-field
(Gruber-Mullins)\cite{Gruber67,Bartelt90} argument, which produces an
expression relating the variance to $\tilde{A}$. In recent years, theories
from two new viewpoints have deduced different relations of
$\tilde{A}$ to the variance of the Gaussian.  More recently, we have
recognized that the TWD might  

\noindent \underline{~~~~~~~~~~~~~~~~~}

\noindent $^*$Corresponding author:
        {\tt einstein@physics.umd.edu}, fax: (+1) 301-314-9465

\noindent $^{\dag}$ Dep't of Physics, Texas A\&M University--Commerce, Commerce, TX 75429
USA

\noindent$^{\ddag}$ LPS Grephe, UJF (CNRS) Grenoble I, B.P. 87, F-38402 St Martin
d'Heres, France

\noindent better be described using a simple expression arising from random
matrix theory, called the ``generalized Wigner surmise." As these results emerged,
they have been published in several different articles
\cite{EP99,Giesen00,RCEG00,beyond}.  The goal of the present paper is to collect
succinctly the important results, to provide a global view of progress on this
problem, to preview forthcoming results \cite{CRE}, and to point out areas where
further progress is needed. 

The following initial comments indicate our guiding philosophy: 1) The continuum
approximation noted above is part of the step continuum approach to vicinal
surfaces.  In this perspective \cite{JW99}, the mesoscopic behavior of the step is
characterized in terms of three parameters: the step stiffness $\tilde{\beta}$, the 
interaction strength $A$ (or its equivalent), and a parameter representing the
dominant kinetics (a kinetic coefficient or diffusion constant times carrier density). Hence, a
knowledge of $\tilde{A}$ is crucial to a proper description. 2) In this approximation, because
step overhangs are physically forbidden, the set of step configurations in 2D space maps into the
world lines describing the evolution of non-crossing particles (spinless fermions or hard bosons)
in 1D space.  This mapping is what leads to most of the progress in theoretical understanding. 
3) In experiments to date, investigators have measured the distribution of terrace widths
$\ell$.  This correlation function in essence is a many-particle correlation function, since one 
measures the probability of finding a pair particles separated by $\ell$ {\it with
none between them}.  (It is much easier for theorists to compute the probability of
finding a pair particles separated by $\ell$, regardless of how many particles are
between them; note that this two-particle correlation function should be equivalent to
the many-particle one for step separations much smaller than the mean separation
$\langle \ell \rangle$.)

In experimental systems (cf.\ Table 2, below),
$\tilde{A}$ is typically between 0 and 15 \cite{Giesen00,RCEG00,JW99}. (Whereas
occasional values up to nearly 4000 have been reported\cite{DZP} for $\tilde{A}$,
our belief is that values above about 20--25 are indicative of anomalous behavior of
some sort.)  Exact theoretical information is available only for
$\tilde{A}=0$ and
$\tilde{A}=2$ \cite{MehtaRanMat,Guhr98}, as well as in the limit $\tilde{A}
\rightarrow \infty$ \cite{f93,f00,SB99}.  Hence, to assess the merits of various
approaches for general $\tilde{A}$, we have generated well-characterized
distributions numerically.  We have
then compared each of the theoretical predictions with these calibration
standards.

In Sec.\ \ref{sec:key} we collect and synthesize the main results first for the
traditional Gaussian analysis of TWDs and  the competing ways of interpreting their
variance in terms of $\tilde{A}$, then for the generalized Wigner distribution arising
from the theory of fluctuating systems. Sec.\ \ref{sec:useful} recounts concisely several
highlights of previous explorations of these ideas. These include useful
results on fitting procedures, an estimate of when discreteness becomes
important, and a procedure to gauge how many {\it independent} measurements are
contained in an image. Sec.\
\ref{sec:expt} gives a brief summary of findings in applications to experimental
data, with an emphasis on trends.  In Sec.\
\ref{sec:new} we present previews of unpublished results concerning new
directions in understanding TWDs with greater insight and in more complicated
situations.  Finally we offer brief conclusions and comments on connections with
other active subjects in condensed matter physics.

\section{Key Results}
\label{sec:key}

\subsection{Gaussian Approximations to TWDs}
\label{ssec:Gaussian}
 
It is convenient and natural to divide $\ell$ by its average value, thus constructing the
dimensionless parameter $s \equiv \ell/\langle \ell
\rangle$.  Then the TWD, $P(s)$, is not just normalized but has unit mean.  
The Gaussian approximation to the TWD is then written: 
\begin{equation}
  P(s) \approx P_{\rm G}(s) \equiv \frac{1}{\sigma_G \sqrt{2\pi}}
              \exp\left[-\frac{(s-1)^2}{2\sigma_G^2}\right] \; . 
\label{e:Ps}
\end{equation}

Gaussians are typically chosen, not just for their simplicity,
but because their use can be justified readily for strong
elastic repulsion between steps.  
In this limit the motion of each step tends to be confined near its 
mean position, a Gruber-Mullins (GM) argument (in which a single step 
is treated as active and its two neighbors are fixed at twice $\langle \ell
\rangle$) shows 
that~\cite{Gruber67,Bartelt90}
\begin{equation}
  \sigma^2 = K_X\tilde{A}^{-1/2},
\label{e:GM}
\end{equation}
where the subscript X anticipates that there will be different
proportionality constants in different approximation schemes, indicated by
X. For the Gruber-Mullins case, with interactions only between nearest-neighbor
steps, $K_{GM(NN)}= 1/\sqrt{48} \approx 0.144$.   For the Gruber-Mullins case, if
all steps are allowed to interact with  $A/\ell^2$, then 48 in $K_{GM(NN)}$ is
replaced by $8\pi^4/15 \approx 52$, decreasing the variance by a scant 3$^+$\%;
i.e., $K_{GM(all)} \approx 0.139$.

The Grenoble group\cite{PM98,IMP98} pointed out recently that the variance in Eq.\
(\ref{e:GM}) using $K_{GM}$ underestimates (for given $\tilde{A}$) the true
variance. Their arguments are based on two ideas. First, the contribution of the
{\em entropic} repulsion {\em decreases}  with increasing energetic repulsion;
physically, large energetic repulsions diminish the chance of neighboring steps 
approaching each other, where the non-crossing condition  underlying the entropic
repulsion becomes significant.  Thus, for very large $\tilde{A}$ the entropy of
interaction becomes negligible, so that the only entropy is that of the individual
steps.  Secondly, if {\em both} steps bounding a terrace fluctuate
independently, then the variance of the TWD should be the {\em sum} of the 
variances of the fluctuations of each step, i.e.\ {\em twice} the variance
obtained in the Gruber-Mullins picture (in which there is  a {\em single}
``active'' step between a pair of straight/rigid neighboring steps).  This
factor is reduced modestly by corrections due to the [anti]correlations
\cite{Masson94} of neighboring steps. As a result, in this perspective the factor
of 48 in $K_{GM(NN)}$ should decrease to 14.80, increasing the variance for a
particular $\tilde{A}$ by a factor of 1.801.
 
Including \underline{\bf e}ntropic repulsions in an \underline{\bf 
a}verage way (mnemonically  denoted $X = EA$, the two highlighted
letters)\cite{EP99} rather than discarding them extends to smaller
$\tilde{A}$ the range of viability of this (modified) asymptotic limit.
Explicitly,
$\tilde{A}$ is replaced in Eq.\ (\ref{e:GM}) by an effective interaction 
strength $\tilde{A}_{\rm eff}$ obtained from the cubic term of the expansion of the 
projected free-energy of a vicinal surface as a function of 
misorientation slope.\cite{JRS} The resulting enhancement is
\begin{equation}
   \! \frac{\tilde{A}_{\rm eff}}{\tilde{A}} \! \equiv \!
\frac{1}{4\tilde{A}}\left(\sqrt{4\tilde{A} + \! 1}+1\right)^2
\! \sim 1 \!+\! \tilde{A}^{-\frac{1}{2}} + \! \frac{\tilde{A}^{-1}}{2}
+  \ldots .
\label{e:Aeff}
\end{equation}
Explicitly, Eq.\ (\ref{e:GM}) 
becomes $\sigma^2 \sim K_{EA}\tilde{A}_{\rm eff}^{-1/2}$, with values for
$K_{EA}$ given in Table 1.  In this case, $K_{EA(NN)}$ is nearly 10\%
larger than $K_{EA(all)}$.\footnote{A value equivalent to $K_{EA(all)}$ = 0.247 was found
explicitly in a calculation using the harmonic, lattice approximation of the
Calogero-Sutherland model \cite{SB99}, as well as implicitly in earlier studies
\cite{f93,KO82,f92}, and seems to be the exact asymptotic coefficient \cite{f00}.}  Since this
approach represents the limit of minimally important entropic interactions, presumably this large
ratio is an upper bound, approached for large
$\tilde{A}$, while the smaller (3$^+$\%) ratio of the GM case is more
appropriate for weaker $\tilde{A}$.  (In the free-fermion limit
($A$ = 0), the NN and ``all"
cases obviously must be the same!)

The preceding approaches make a continuum approximation 
along the ``time-like''
$\hat{y}$-direction but maintain discrete steps.  By making a continuum
approximation in the
$x$-direction as well  and invoking correlation functions from \underline{r}oughening
theory (so denoted $X=R$),  the Saclay
group\cite{Masson94,Barbier96,LeGoff99} arrived at a result of the form of
Eq.\ (\ref{e:GM}), again with  $\tilde{A}_{\rm eff}$ replacing $\tilde{A}$,
in which
$K_R \! = \! 2/\pi^2 \approx 0.203$.  

  Since the various Gaussian approaches make different fundamental
approximations, the detailed relationships between the width of
the Gaussian and 
$\tilde{A}$ differ notably.  Even when a
TWD can be well fit by a Gaussian,the estimation of
$\tilde{A}$ can be ambiguous.


\subsection{Symmetry and Wigner Approximation to TWDs: Continuous
Generalized Wigner Distribution}
\label{ssec:Wigner}

In considering high-lying energy levels in nuclei, Wigner long ago proposed that
fluctuations in their spacings in energy should exhibit certain universal features
depending only on the symmetry---orthogonal, unitary, or symplectic---of the
couplings.  This work, embedded in random-matrix theory\cite{MehtaRanMat,Guhr98},
has had profound and widespread implications for characterizing a wide range of
fluctuation phenomena \cite{Guhr98,SB99}. Since TWDs are an example of equilibrium
fluctuations \cite{EP99,Giesen00,RCEG00}, this body of
knowledge should be applicable to them.  The explicit connection is based on the description of
steps using the (Calogero\cite{C69}-)Sutherland
\cite{SutherGaus71,SutherCirc71} model of spinless fermions on a large ring (essentially 1D
with periodic boundary conditions) interacting with a repulsion decaying as the
inverse square of separation.  Remarkably, the distribution of
interparticle spacings along the ring (i.e. the TWD) is equivalent to the
distribution of the above-mentioned energy spacings, which can be solved exactly
by random-matrix methods for the three symmetries \footnote{By construction, the
wavefunction $\Psi$ of the Sutherland model is the product of the $\varrho$/2 power of
the differences of all pairs of particle positions, so that $|\Psi^2|$ was recognized to
be identical to the joint probability density function for the eigenvalues of random
matrices from a Gaussian \cite{SutherGaus71} or a circular \cite{SutherCirc71}
ensemble.} According to the so-called Wigner surmise, these three exact solutions for
the distribution of fluctuations can be approximated by\cite{EP99}
\begin{equation}
  \label{e:Wigner}
   P_\varrho(s) = 
    a_\varrho s^{\varrho} \exp \left(-b_\varrho s^2\right) \, .
\end{equation}
\noindent The three symmetries correspond to the values
$\varrho =$ 1, 2, or 4, respectively. The constants
$b_\varrho$ (associated
with unit mean of $P(s)$) and $a_\varrho$ (deriving from normalization) are
\begin{equation}
b_\varrho = 
  \left[\frac{\Gamma \left(\frac{\varrho +2}{2}\right)}
             {\Gamma \left(\frac{\varrho +1}{2}\right)}\right]^2  \quad\mbox{and}\quad
    a_\varrho = \frac{2b_\varrho^{(\varrho + 1)/2}}
	          {\Gamma \left(\frac{\varrho+1}{2}\right)}.
\label{e:abr}
\end{equation}

\noindent This surmise was used to describe the spacings not of particles in real space but
rather of energy levels, first in nuclei, later in chaotic systems \cite{Haake91}.

The variance of $P_\varrho(s)$ is just 
\begin{equation}
\label{e:sra}
   \sigma^2_W   =  \frac{\varrho + 1}{2b_\varrho} - 1 \;
\lower1.ex \hbox{$\displaystyle \sim \atop \varrho \to \infty$} \;
\frac{1}{2\varrho}.
\end{equation}

\noindent  The approximations prove to be
outstanding, accurate to better than $\pm 0.004$ for the latter two cases
(cf. Fig.~1 and Ref.\ \cite{Haake91}, Fig.\ 4.2a).  From the mapping of the
step problem onto the Sutherland Hamiltonian\cite{SutherGaus71} comes the relation

\begin{equation}
\tilde{A} = \varrho(\varrho-2)/4,
\label{e:Arho}
\end{equation}

\noindent (By inverting
Eq.~(\ref{e:Arho}) to obtain $\varrho$ as a function of $\tilde{A}$, we can make
the important and useful identification of $\varrho$ as
2$\surd\tilde{A}_{\rm eff}$, as given in Eq.~(\ref{e:Aeff})!) 

For the three special values of
$\varrho$, Eq.~\ref{e:Wigner} accounts for the cases $\tilde{A} =$ --1/4, 0, or 2,
respectively.  The value 0 corresponds to steps interacting only via the entropic
repulsion, whereas the negative value corresponds to an attraction, which cannot be
produced by the generic elastic interaction between steps (except perhaps in
abnormal cases in which there are strong in-plane dipoles at the step
edges \cite{JRS,Bonzel00}). 
The third case, $\tilde{A}$ = 2, corresponds to a rather moderate
repulsion.  As documented in Table 1, the variance of Wigner's $P_\varrho(s)$ is nearly
the same as the exact value.  The Saclay and the GM estimates are a few percent
too low, while the modified Grenoble estimate 
is much too high. 

The crucial question is what to do for more general values of $\tilde{A}$.  We simply
use Eq.~(\ref{e:Wigner}) for {\it arbitrary} value of
$\varrho \ge 2$, with $\varrho$ related to $\tilde{A}$ by Eq.~(\ref{e:Arho}). 
For brevity, we refer hereafter to this distribution, for general $\varrho$, as
the CGWD (continuum generalized Wigner distribution). In 

\begin{figure}[b]
\centerline{\psfig{file=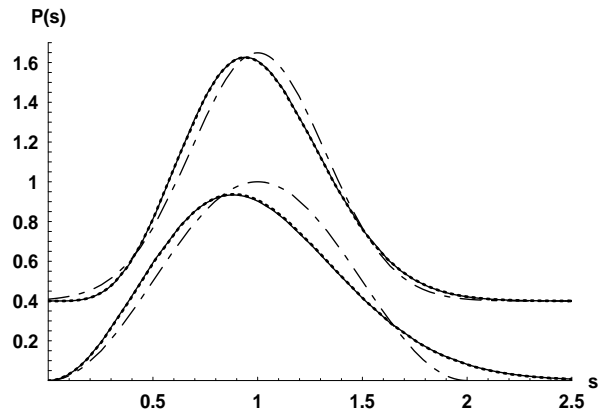,width=8cm,height=6cm,angle=0}}
\caption[shrt]{
$P(s)$ vs.\ $s \equiv \ell/\langle \ell \rangle$ for the [sixth approximant \cite{Joos91} to
the] exact ``free-fermion", $\tilde{A}$=0 result (solid curve), the Gruber-Mullins
approximation $\sin^2(\pi s/2)$(long-short dashed curve), and the
$\varrho$=2 Wigner surmise result (dotted curve), barely distinguishable from the
exact result).  Offset upward by 0.4 for clarity, a similar plot of an approximant
of the exact result for $\tilde{A}$=2 \cite{Joos91}, the Gruber-Mullins Gaussian
approximation
$(24/\pi^2)^{1/4}\exp(-\surd 24 (s-1)^2)$(long-short dashed curve), and the
$\varrho$=4 Wigner surmise result (dotted curve).}
\end{figure}

\noindent contrast to the three
special cases, there are no symmetry arguments to justify the CGWD form. 
We offer the following arguments in its support, although ultimately one must
rely on numerical checks. 

\noindent 1) It seems plausible that $P_{\varrho}(s)$ is a
decent approximation of the TWD for values of
$\varrho$ between 2 and 4 since the range in parameter space is small.  In any
case, the arguments supporting the approaches leading to any of the Gaussian
approximations fail in this regime.

\noindent  2) Extrapolation of the CGWD to values of
$\varrho$ greater---possibly much greater---than 4 is of more concern.   For very
large $\tilde{A}$, the argument underlying the Grenoble viewpoint becomes
compelling.  In this limit, the leading term in the
expansion of $\sigma^2_W$ in Eq.~(\ref{e:sra}) implies that 
$K_W = 1/4$ in Eq.~(\ref{e:GM}), with $\tilde{A}_{\rm eff}$ replacing
$\tilde{A}$. Thus, as listed in Table 1, the CGWD variance approaches the
[modified] Grenoble estimate nicely, while the Saclay estimate is notably
too small.  Since the CGWD
does well in the limit of very large
$\varrho$ as well as at $\varrho$=4, it is a promising candidate for an
interpolation method between these values. 

\noindent  3) As a function of $s$, the CGWD not only has the Gaussian behavior
expected (based on analogies with random walkers) at large step separations but
also reproduces the exact power of $s$ for $s \ll 1$:  In this limit, the
many-step correlation function becomes identical to the pair correlation
function, due to the vanishing probability of any other step lying between the
pair of steps separated by
$s$. Several workers have shown that in this limit, the pair correlation
function is proportional to $s^{\varrho}$, with a prefactor similar (within at
least a few percent for physical values of $\varrho$) to
$a_{\varrho}$\cite{f93,f92,h95,lps95,h94}. 

\noindent 4) We can derive the CGWD from a Schr\"odinger-equation
approach \cite{beyond}, as discussed below in Section \ref{sec:beyond}.  This approach
has the further benefit of allowing one to consider more general potentials than the
asymptotic form of the elastic repulsion. 

\subsection{Preliminary Numerical Results}

To test numerically the accuracy of Eq.~(\ref{e:Wigner}) we apply standard Monte Carlo
methods to the most elementary model that contains the necessary physics, the
terrace-step-kink (TSK)  model. In the TSK model the only thermal
excitations are kinks of energy $\epsilon$ along the steps.  The stiffness
$\tilde{\beta}_{TSK}$ of an isolated step---needed to extract $A$ from
$\tilde{A}$---is simply
$2k_BT(a_{\parallel}/a^2_{\perp})\sinh^2(\epsilon/2k_BT)$ \cite{BEW92}. Here
$a_{\parallel}$ is the unit spacing along a step edge ($\hat{y}$), and $a_{\perp}$
is the $\hat{x}$ component of a kink.  This model is obviously
{\it discrete} in the $\hat{y}$ as well as the $\hat{x}$
directions \cite{Bartelt90,Masson94,LeGoff99,BEW92}. For simplicity we consider a
vicinal simple cubic lattice with unit lattice constant:
$a_{\parallel} = a_{\perp} = 1$.  Periodic boundary conditions are imposed in both
directions.  To minimize finite-size effects, the length of the system in the
$\hat{y}$-direction,
$L_y$, should be substantially larger than the characteristic distance $y_{\rm coll}$
along $\hat{y}$ between close approaches of adjacent steps: $ \langle \ell
\rangle^2 \tilde{\beta}/k_BT$ \cite{BEW92}.  The choice of the mean spacing
between steps requires particular care.  We shall show below that if
$\langle \ell \rangle$ is 4 or smaller, finite size effects may contaminate the
results extracted from the CGWD (since it is based on a continuum approximation).
On the other hand the minimum acceptable value of $L_y$ increases like $\langle
\ell \rangle^2$.  Furthermore, too low a temperature results in slow dynamics
and a high stiffness, making demands on $L_y$, while too high a temperature
leads excessive step wandering and breakdown of the approximations underlying
the viability of the TSK model.  We are preparing a careful discussion of these
considerations \cite{CRE}, which includes transfer-matrix calculations in
addition to Monte Carlo simulations.  

In Fig.~2 we provide some preliminary results for the case  $\langle
\ell \rangle = 6$ at $k_BT/\epsilon = 0.5$, with $L_y = 200$ and the number of steps $N
= 10$.  Data from runs using the standard Metropolis algorithm\cite{Metro53} were taken,
after equilibration, over 10$^5$ Monte Carlo steps per site.  In addition we use the
``refusal-free'' $n$-fold way
\cite{Bortz75,NovCIP}, especially for large
$\tilde{A}$ (or at low $T$).  There it is much more efficient than the Metropolis
algorithm, which requires many attempts before making a change. The elastic repulsion is
here considered only between neighboring steps, a common simplification in Monte Carlo
\cite{Bartelt90,Barbier96}, with the accordant modest underestimate of
$\sigma^2$ noted in Subsection A. (In Ref.~\cite{CRE} we will also extend the
inverse-square repulsions to further neighbors.) Our algorithm includes ``corner
exclusion" in addition to standard edge exclusion, based 

\begin{figure}[b]
\centerline{\psfig{file=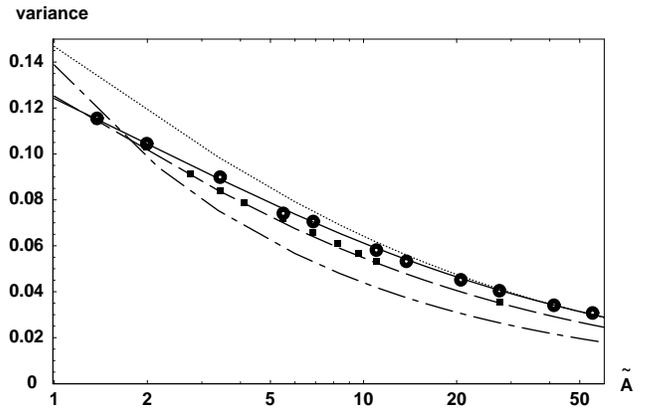,width=8cm,height=6cm,angle=0}}
\caption[shrt]{
Plot of the variance $\sigma^2$ as a function of $\tilde{A}$ on a logarithmic scale.
plotted for the CGWD [``Wigner distribution"] (solid curve), 
the modified Grenoble (dotted curve) 
and Saclay (long dashed curve) Gaussian distributions, and the Gruber-Mullins Gaussian
approximation (long-short dashed curve).  The CGWD curve passes essentially
directly over the exact value of the variance at $\tilde{A}$=2.  Monte Carlo data
generated using the Metropolis algorithm are  depicted by
$\Box$'s; data produced with the
$n$-fold way algorithm are shown as $\bullet$'s.}  
\end{figure}

\noindent on some evidence that it
provides the better discrete analogue of the continuum model; the consequent modest
upward shift of $\sigma^2$ is in the opposite direction of that due to restriction to
nearest-neighbor step-step repulsions. (See Ref.~\cite{CRE} for more details.)

Along with the numerical results, the various
predictions of the variance are plotted as functions of the physical variable
$\tilde{A}$.  A logarithmic scale is used for the horizontal axis so as not to give
undue visual emphasis to larger values of
$\tilde{A}$ nor to blur the region of rapid variation for small $\tilde{A}$, for
which an exact calibration point exists. The physical values of
$\tilde{A}$ range from near 0 up to the mid teens.  A few larger values have
been reported, but there are suspicions that more than simple elastic repulsions
are involved.  There are relatively few reports of small but non-zero values of
$\tilde{A}$.  We suspect that one reason is that any of the Gaussian
approximations manifestly fail in this regime because the distribution becomes
strongly skewed.\footnote{However, the idea of estimating $\tilde{A}$ using this
skewness \cite{EP99} did not prove to be fruitful when confronting experimental data
\cite{Giesen00}.} Before the recognition of the utility of the Wigner distribution,
one could not deal quantitatively with small $\tilde{A}$ \cite{SBV}.

\section{Useful Results for Interpreting Experiments}
\label{sec:useful}

\subsection{Extracting $\tilde{A}$ from Variance}
If one accepts the CGWD as the optimal way to analyze TWDs, then Eq.~(\ref{e:sra}) shows
how to estimate the variance from
$\tilde{A}$.  However, experimentalists usually seek the reverse. An excellent 
estimate \cite{RCEG00} of
$\tilde{A}_{\rm W}$ from the variance can be derived by expanding $\sigma^2_W$ as given
in  Eq.~(\ref{e:sra}) in powers of $\varrho^{-1}$.  This series can then
be reverted to give
$\varrho$ as a function of $\sigma^2$ \cite{RCEG00}.  Then using
Eq.~(\ref{e:Arho}) gives the estimate
\begin{equation}
   \tilde{A}_W
    \approx  \frac{ 1}{16} \left[(\sigma^2)^{-2}
          - 7 (\sigma^2)^{-1}
          + \frac{27}{4}
          + \frac{35}{6} \sigma^2\right]  \, ,
\label{e:as}
\end{equation}
\noindent with all four terms needed to provide a good approximation
over the full physical range of $\tilde{A}$.  The Gaussian methods described
earlier essentially use just the first term of this expression and adjust the
prefactor.  When $\tilde{A}$ is not weak (see Ref.~\onlinecite{RCEG00} for
explicit guidelines.), those who for some reason prefer not to use
Eq.~(\ref{e:Wigner}) to gauge $\varrho$ (and thence $\tilde{A}$) can extract the
variance from a Gaussian fit and then applying Eq.\ (\ref{e:as}) is a reasonable
procedure.
 
When dealing with tabulations of data analyzed in the traditional way
\cite{JW99}, i.e. using the inverse of Eq.~(\ref{e:GM}) with $X=GM(NN)$, it is
useful to recast Eq.~(\ref{e:as}) in a form that indicates the factor by which
the estimate $\tilde{A}_W$ based on CGWD exceeds the traditional estimate
$\tilde{A}_{GM(NN)}$ (denoted $\tilde{A}_G$ for brevity):
\begin{equation}
  \label{e:aa}
  \tilde{A}_W/\tilde{A}_G \: [\equiv A_W/A_G] \:
    \approx  3  -  21\sigma^2 +\frac{81}{4}\sigma^4 +\frac{35}{2}\sigma^6\, .
\end{equation}

\noindent As noted parenthetically, Eq.~(\ref{e:defAtilde}) implies that the
ratio of the physical interaction strengths is the same as that of the
dimensionless strengths.
Since $A \propto \tilde{A}$ we can use this relation in Table 2 to update 
tabulated $A$'s in Ref.~\cite{JW99}.

\subsection{Gaussian Fits of the Generalized Wigner Distribution}
\label{sec:gausswigner}

Since TWDs for strong repulsions are well described by Gaussians, one
expects---and finds---that the CGWD should be well approximated by a Gaussian in
this limit.  In Ref.\ \cite{RCEG00} a quantitative assessment is given of how
closely the two distributions correspond as a function of $\varrho$.
At the calibration point (for which an exact
solution exists) for repulsive
interactions  ($\varrho =4$),  the relative difference of the standard deviation
of a Gaussian fitted to $P_{\varrho}(s)$ from the actual standard deviation of
this CGWD (viz. the square root of the second moment of $P_{\varrho}(s)$ about its
mean of unity) is  around 1\%,
and decreases monotonically with increasing
$\varrho$.  For  this range ($\varrho \geq 4$) differences between estimates of
$\tilde{A}$ obtained from CGWD and the various Gaussian fit
methods are predominantly  due to different philosophies of
extracting $\tilde{A}$ from
$\sigma$ rather than from differences in the fitting methods.

In contrast to the Gaussian approximations, the peak of the CGWD must perforce
(due to unit mean) lie below one.  Specifically, for $\tilde{A}$ = 0 and 2, the
maximum of $P_{\varrho}(s)$ occurs at $s$ = 0.886 and 0.940, respectively, while
the limiting value for strong repulsions is 
$1 \! - \! 0.125 /\! \surd \tilde{A}_{\rm eff}$ \cite{EP99}.  Formulas have been
derived \cite{RCEG00} indicating the errors in fitting $\varrho$ due to
errors in the first or zeroth moment of the distribution.

\subsection{Wigner Distribution as a 2-Parameter Fit}
\label{ssec:wigner2param}

In applications to experimental TWDs, the CGWDs giving the best fits sometimes
have  first moments that differ somewhat from the first moments 
of the data, especially in cases termed ``poor data" \cite{Giesen00,RCEG00}
which exhibit a small  ``hump'' at large values of $s$, beyond the peak near
unity [see Section \ref{sec:expt} below].  
Moreover, it can be desirable to 
determine the scaling length (the ``effective mean,'' which 
equals the first moment for ideal CGWDs) and the
variance in a single fitting procedure rather than to predetermine this length
from the first moment.  
This ``refined" scaling implies that the argument of $P_{\varrho}$ should be
$\ell/\bar{\ell}$, where $\bar{\ell}$ denotes the characteristic length
determined  along with $\varrho$ in a two-parameter least-squares fit of the data
to a CGWD.  Since $s$ is still determined from the raw data as 
$\ell/\langle \ell \rangle$, the refined scaling translates into replacing
$s$ by $s \langle \ell \rangle/\bar{\ell}$ in the argument of the distribution. 
If the integration variable $s$ were similarly replaced, then the refined
scaling would amount to a redefinition of a dummy variable, and normalization
would still be realized.  Since the independent variable is kept as
$s$, we make the replacement:

\begin{equation}
P_{\varrho}(s) \rightarrow (\langle \ell \rangle/\bar{\ell}) P_{\varrho}(s
\langle \ell \rangle/\bar{\ell}) \quad\mbox{i.e.}\quad 
(\langle \ell\rangle/\bar{\ell}) P_{\varrho}(\ell/\bar{\ell})
\label{e:Psa}
\end{equation}

\noindent In the specific applications to data in subsections A and B of the next
section, $\langle \ell \rangle/\bar{\ell}$ tends to be
greater than unity, typically by several percent, but it is unclear whether this
is true for semiconductors or other metals.
In our companion Monte Carlo simulations \cite{CRE}, where we have greater
control of purity and uniformity than in experiments, the optimal $\bar{\ell}$ is
essentially identical to $\langle \ell \rangle$: there is no need for the added
flexibility of the two-parameter fit.

\subsection{Effects of Lattice Discreteness}

For actual crystals as well as for the TSK model used in numerical simulations, the
variable $s$ cannot assume a continuum of values as implicitly assumed in writing
Eqs.~(\ref{e:Wigner},\ref{e:abr}); the only possible values of $\ell$ are integer
multiples of $a_{\perp}$.  If this restriction is
placed on the values of $\ell$ used to generate the arguments $s$ of Eq.~(\ref{e:Wigner}),
then we have constructed a {\it discrete} generalized Wigner distribution
(DGWD).\footnote{Discreteness also introduces the possibility of a roughening transition
from a vicinal to a high-index-facet surface \cite{LeGoff99,SZS98}.  This interesting
phenomenon does not seem to play a role in the physical systems under study here.} We
use the same value of $b_{\varrho}$ as in Eq.~(\ref{e:abr}), even though it is no longer
guaranteed to produce unit mean (or the same variance) as it does for the CGWD. Since
these vicinals are technically rough, there is no need for
$\langle \ell \rangle$ to be an integer multiple of $a_{\perp}$ (or otherwise in
registry with the terrace plane), though it is common to make this choice in
simulations.  

Scaling of discrete TWDs for the free-fermion case ($\tilde{A}=0$) was
demonstrated nearly a decade ago \cite{Joos91}.  Inspired thereby, we
\cite{RCEG00} explored the effects of discreteness, first choosing values of $\langle
\ell \rangle$ and
$\varrho$ to specify a DGWD, then numerically performing two-parameter fits using
CGWD formulae [Eqs.~(\ref{e:Wigner}), (\ref{e:abr}), (\ref{e:Psa})] to produce estimates
of $\varrho_c$ and (via Eq.~(\ref{e:Arho}))
$\tilde{A}_c$ .

Among many minor observations, two major themes stand out: 
First, $\langle \ell \rangle \! \geq \! 4$,
$\tilde{A}_c$ provides a reasonable estimate of
$\tilde{A}$ over the range of physically reasonable
dimensionless repulsions.  Furthermore, at fixed values of
$\tilde{A}$ the error
in $\tilde{A}_c$ diminishes as $\langle \ell \rangle$ increases.
Second, as the TWD
becomes narrower (i.e.\ for sufficiently large $\tilde{A}$ or
$\varrho$), $\tilde{A}_c$ becomes a questionable estimate for
$\tilde{A}$; study of the cases $\langle \ell
\rangle/a_{\perp}$ = 2--6 suggests that this breakdown occurs for $\varrho$
near $(\langle \ell \rangle/a_{\perp})^2$.  This threshold corresponds to the squared
interstep spacing being comparable to the variance.  

For very large $\tilde{A}$, seemingly just above the range of greatest physical
significance, there are more general indications of the breaking down of the
continuum approximation.  E.g., for $\tilde{A}$ in the upper teens, there begin to be
ambiguities in the application of Eq.~(\ref{e:Arho}) \cite{CRE}.  With periodic
boundary conditions one can still get elementary excitations that are extended along
a step (i.e. along
$\hat{y}$), so long as they are ``in phase" in $\hat{x}$, but with more realistic
conditions (with various sorts of defects hindering the fluctuations of occasional
steps) , the elementary excitation becomes individual ``teeth" (kink-antikink pairs
separated by one spacing along
$\hat{y}$)
\cite{HvB}.  Then the idea of step stiffness also breaks down, and with it the
concept of
$\tilde{A}$ (see Eq.~(\ref{e:defAtilde})).

The main implication is that analyses of highly misoriented vicinal surfaces with CGWD
should be viewed with caution.  E.g.\ the (1,1,7) for close-packed
steps on surfaces vicinal to
\{1 0 0\} planes of fcc crystals corresponds to $\langle \ell \rangle = 3$.  
For \{1 1 1\} fcc surfaces, the corresponding Miller indices are (5 3 3) for A steps 
(\{1 0 0\} microfacets) and (2 2 1) for B steps (\{1 1 1\} microfacets)\cite{EE93}. 

The obstacles posed by discreteness are not vagaries of Wigner
distributions.  High misorientation causes similar problems when the mean and
variance
of discretized Gaussian TWDs are analyzed as though they were continuous
Gaussian functions. (See Ref.\ \cite{RCEG00} for details.)

\subsection{Estimate of Number of Independent Measurements}
\label{sec:sample}

In order to estimate uncertainties in the determination of the TWD and, ultimately,
$\tilde{A}$, it is important to have a realistic value of the  number of {\it
independent} measurements, a number generally 
much smaller than the total number of measurements.
To make a rough estimate, one can compute the correlation 
function\cite{StatisticsForExperimenters} of the terrace widths $\ell_n(y)$ between 
steps $n$ and $n\! + \! 1$:
\begin{eqnarray}
   C_n(y)  =   \frac{
         \frac{\sum_{n^\prime}^{N-n}\sum_{y^\prime=1}^{L_y-y}
                    \ell_{n^\prime}(y^\prime) \ell_{n^\prime+n}(y^\prime+y)}
                   {(N-n)(L_y-y)}
 - \langle \ell \rangle^2}{\langle \ell^2 \rangle - \langle \ell \rangle^2}
\end{eqnarray}
is calculated, where $N$ is the number of terraces in the image.
The correlation function along the steps decays exponentially
as $C_0(y) \sim \exp(-y/\xi_y)$, 
where $\xi_y$ is proportional to $y_{\rm coll}$ (cf.\ Eqs.~(5), (12), and (26) of
Ref.~\cite{BEW92}), but can be measured directly. 
The correlation function between steps 
is more complicated.  As noted in Table 1, $C_1(0)$ is negative\cite{MehtaRanMat}; 
$|C_n(0)|$ tends to decrease rapidly with increasing $n$. 
Setting $c$ as a small cutoff ($c \! = \! 0.1$ is recommended \cite{RCEG00}), we determine
$y_c$, the smallest value of $y$ for which 
$|C_0(y)| \leq c$ when $y \geq y_c$, 
and $n_c$, the smallest $n$ so that
$|C_n(0)| \leq c$ for all  $n \geq n_c$.
Then the number of ``independent'' terrace widths will be approximately 
$(L_y/y_c)(N/n_c)$ rather than $L_yN$, as might be naively guessed.  A rough test
calculation \cite{RCEG00} shows that the reduction factor can be nearly two orders of
magnitude, emphasizing the need for using several STM images to obtain decent
statistics.

\section{Applications to Experimental TWDs}
\label{sec:expt}

In two papers \cite{Giesen00,RCEG00}, we made extensive applications of the ideas
presented above to M. Giesen's voluminous data on vicinal Cu \{100\} and \{111\}
surfaces, each at three different misorientations, and these six cases at various
temperatures.  In all, around 30 different cases were considered.  In addition, our
ideas were tested successfully on data for vicinal Pt(110), which has a small
$\tilde{A}$ and so is not amenable to the Gaussian approaches used heretofore.  The
purpose of this section is to summarize the tabulations and discussions in those
papers. 

\subsection{Copper: Moderately Strong Repulsions}

The Cu TWDs can be sorted into three groups based on a visual assessment of their quality
\cite{Giesen00,RCEG00}:  A ``good'' TWD changes height essentially monotonically below 
	the peak and again above it; there is minimal scatter in the data points. 
 An ``OK'' TWD has
more scatter, with small dips and peaks 
	introduced by variations (within the limits of the 
	general margin of error) of single data points. 
 A ``poor'' TWD has a double-peak or hump at large $s$; 
	correspondingly, the position of the (main) peak occurs 
	noticeably below $s$ = 1, even when the peak is fairly 
	narrow and the skewness minimal.  
	The judgment that this data is ``poor'' is based both on 
	the intuition of the {\em experimenter} and on the following 
	argument: 
        A second peak at 
	large $s$ would be characteristic of the onset of faceting; 
	however, ``poor'' data tends to occur at high temperatures,
	whereas faceting should be more important at low temperatures. 

The data fits exhibited several general trends.  In almost all cases,
the value of $\bar{\ell}$ derived from the two-parameter fit to a CGWD is
smaller
than $\langle \ell \rangle$ given by the mean
of the TWD (and the opposite shift in the exceptional cases is very small); likewise, the
directly measured values of the variance are almost always larger than the values obtained by
any of the three fitted curves (cf.\ Sec.~7 of Ref.\ \onlinecite{RCEG00}).
The value of
$\varrho$ is higher for the two-parameter CGWD fit than for the single-parameter version,
and the associated value of
$\sigma^2$ typically closer to that deduced from the Gaussian fit.  For
``good'' data, $\bar{\ell}/\langle \ell \rangle$ differs from unity by a few percent, and the
change in
$\varrho$ and $\sigma^2$ is negligible.  For ``poor" data, $\bar{\ell}/\langle \ell \rangle$
is at least twice as far below unity, and the two-parameter-fit curve is narrower than the
single-parameter-fit curve.   The tails or humps in
the experimental TWDs seem to be responsible  for the systematic discrepancies
in the fits, especially the smaller mean and smaller variance of the fits
relative to the direct measurements.  

A remarkable consistency check was obtained for Cu (1 1 13) \cite{Giesen00}.  For
a dozen values of temperature, $(k_BT)^2 \tilde{A}$ was plotted against $T$. 
Since $A$ is expected to be relatively insensitive to thermal change,
Eq.~(\ref{e:defAtilde}) predicts that the plotted curve should decrease like the
stiffness.  To within error bars, such behavior is found, where the stiffness is
computed using an independently determined kink energy.

\subsection{Platinum: Weak Repulsions}

On vicinal Pt(110)
at room temperature, the terraces are 
($1 \! \times \! 2$)
reconstructed, and the steps correspond to 3-unit ``($1\! \times \! 3$)" segments. 
Recent measurements show that the
interaction between their steps is small \cite{SBV}, rendering Gaussian approximations
invalid.  Fits to the CGWD yield $\varrho$ = 2.06 ($\tilde{A}$ = 0.0309) or, when done in
the two-parameter way, $\varrho$ = 2.24 ($\tilde{A}$ = 0.134) \cite{RCEG00}; 
in the latter case, the optimal $\bar{\ell}/\langle \ell \rangle$ is 91\% and the fit is
notably better.
The presence of a high-$s$ bulge indicates this feature is not peculiar to the vicinal
Cu systems of GE.

\subsection{Other Systems}

Additionally, in Table 2 we list the variances measured for several
different experimental systems, along with the value of $\tilde{A}$ deduced from the
CGWD distribution via Eq.~(\ref{e:as}).  The primary goal of this table is to
display general trends in physical systems rather than to provide a comprehensive
account of experiments to date.  As asserted earlier, values of $\tilde{A}$ are
generally below the mid-teens.  On the other hand, $A$ ranges over orders of
magnitude.  If one accepts that the CGWD provides a good accounting in general for
$\tilde{A}$ as a function of the measured variance, then the column labeled
$A_W$/$A_G$ shows that for most systems, the underestimate by using the
Gruber-Mullins approximation is roughly half that of the asymptotic limit.

\section{New Directions}
\label{sec:new}

\subsection{Multistep Distributions}

Experiments to date have focused exclusively on the TWDs, ignoring the possibility of
extracting the distributions of the distances between pairs of steps having $n$ steps,
$n=$ 1,2, or more, between them.  This supplementary data could provide a valuable
consistency check.  For the three special cases $\varrho$=1,2,4, these distributions have
recently been investigated theoretically in a different context \cite{AS99}.  If in
Eq.~(\ref{e:Wigner}) we make the redefinition
$s\equiv(\ell_1 + \ldots + \ell_{n+1})/\langle\ell\rangle$  ($\ell$ being
the terrace width), then this CGWD expression gives a good approximation of the
multistep distribution, provided that the power-law exponent $\varrho$ is replaced by
\begin{equation}
\varrho_n = n +\frac{(n+1)(n+2)}{2}\varrho,
\label{e:AS}
\end{equation}
where $\varrho$ (or, equivalently, $\varrho_0$) is the exponent for the
[single] TWD.
The new constant
$b_{\varrho_n}$ is determined by the condition that the first moment of
$P_\varrho(n,s)$ is $n+1$; besides replacing $\varrho$ by $\varrho_n$ in the
$\Gamma$-function arguments in Eq.~\ref{e:abr}, a factor of $n$+1 must be included in the
denominator.
The normalization constant $a_{\varrho_n}$ can be obtained simply by using
$b_{\varrho_n}$ and $\varrho_n$ in the expression for $a_{\varrho}$ in
Eq.~(\ref{e:abr}).  As for TWDs, these results can be taken to apply to general values
of $\varrho$.  

Preliminary checks using Monte Carlo simulations of the TSK model,
described above, find fine agreement with this multistep CGWD for the
double-terrace-width ($n$=1) case, but just adequate agreement for the case
$n$=2.  Moreover, Table 1 shows that the variance of the sum over the widths of two adjacent
terraces predicted by $P_\varrho(1,s)$ does not display the spectacular agreement with exact
results seen for simple TWDs, viz.\ $n$=0. We suspect that the agreement will further degrade as
more widths are included (higher $n$'s considered), due to weakness in the main assumption in the
derivation of Eq.~(\ref{e:AS}): that the conditional probability density of occurrence of a step
at a given distance from a fixed step, with $n$ steps in between, can be expressed in terms of
the ($n$+1)th power of the corresponding probability for this distance with no intermediate steps.

\subsection{CGWD and Beyond via Schr\"odinger Equations} 
\label{sec:beyond} 
As presented, the
CGWD is formally justified only for the three special values of $\varrho$. Accordingly, we have
developed arguments using Schr\"odinger equations to show that it can be expected to have the
more general validity assumed above \cite{beyond}. The formalism also allows treatment of more
general potentials than the inverse-square term characterizing the long-range behavior of elastic
interactions. Of particular physical importance are the higher-order terms that enter at smaller
terrace widths and an oscillatory interaction mediated by electronic surface states. 

We begin by
defining a wave function $\psi_0(s)$ such that $\psi_0^2(s) \! \equiv \! P_{\varrho}(s)$.
Differentiating twice and using Eq.~(\ref{e:Arho}), we find 
\begin{equation} 
\label{e:diff2}
\! -\frac{{\rm d}^2} {{\rm d}s^2} \psi_0(s) + \left[\tilde{A} s^{-2} -b_\varrho (\varrho+1) +
b_\varrho^2 s^2\right]\psi_0(s) = 0. 
\end{equation} 
The term $\tilde{V}(s) \! = \!
\tilde{A} s^{-2}$ is the dimensionless step interaction. The term $\tilde{U}(s) \! = \!
b_\varrho^2 s^2$ is a dimensionless projected free energy representing interactions with all the
other steps not explicitly considered. Clearly Eq.~(\ref{e:diff2}) can be understood as a
Schr\"odinger equation, with $\psi_0(s)$ the (real) ground state wave function and with
$b_\varrho (\varrho+1)$ the associated eigenvalue. (To consider perturbations from pure inverse
square interactions, one can generate all the eigenfunctions $\psi_n(s)$ of Eq.~(\ref{e:diff2}),
which can be expressed as special functions (see Ref.~\cite{beyond} for details).) In this
framework, by substituting more general potentials for $\tilde{A} s^{-2}$ in Eq.~(\ref{e:diff2})
and solving for the ground-state wavefunction, we can contend analytically with more complicated
potentials. Successful tests are described in Ref.~\cite{beyond}. 

It is tantalyzing to invert the preceding approach to deduce the
underlying interaction potential from the experimental TWD. Since naive implementations prove to
be dangerous, the recommended procedure \cite{beyond} requires considerable computation:
Initially, a parametric approximant of the unknown potential should be constructed using all
available information; crude initial estimates must be made of the values of the parameters. If
the tail of the experimental TWD is Gaussian, a monotonically decaying $\tilde{V}(s)$ and a
quadratic $\tilde{U}(s)$ are anticipated. (If the tail is exponential, $\tilde{U}(s)$ is linear
in $s$, and $\tilde{V}(s)$ can be nonmonotonic.) Then the choice of parameters is optimized by
iteratively minimizing the least-squares difference of the experimental TWD and $|\psi_0(s)|^2$,
where $\psi_0(s)$ is the numerical solution of Eq.~(\ref{e:diff2}) or its equivalent. In a test
of this procedure, for very large or very small $s$ the derived potential $\tilde{V}(s)$ was
significantly different from the potential used to generate the ``experimental" TWD, but they
were quite close to each other over the range where the TWD is large, $0.5 \! \leq \! s \! \leq
1.5$. Hence, to improve estimates of the potential over a large range of $\ell$, one should fit
TWDs measured for several different misorientations (and, if possible, for different
temperatures). 

Applications to vicinal Cu surfaces is in progress \cite{beyond2}. Preliminary
fits can account for secondary humps mentioned in Section \ref{sec:expt} by invoking
nonmonotonically decaying interactions and exponentially decaying tails. 

\subsection{Oscillatory Interactions Mediated by Surface Electronic States}

The preceding subsection began with allusions to oscillatory
electronic indirect interactions between steps \cite{E96}.  When mediated by bulk
electronic states, such interactions between atoms on surfaces decay rapidly with
separation, but if mediated by circularly-symmetric surface states, the envelope of the
oscillatory interaction has the same inverse-square behavior as the monotonic elastic, dipolar,
and entropic repulsions:

\begin{equation}
  \ell^{-2} \cos(2k_F\ell +\phi),
  \label{e:indir}
\end{equation}

\noindent where $k_F$ is the wavevector of the surface state at the Fermi level
\footnote{For non-circular states, the value of the wave vector in Eq.~(\ref{e:indir}) is
that along the Fermi ``surface" at which the electron {\em velocity} is in the
$\hat{x}$ direction; see Ref.~\cite{E96} for details.} and $\phi$ is a phase shift associated
with scattering from the pair of steps.

Consistent behavior was seen in measurements of TWDs on
vicinal Ag(110), including the presence of a surface state in the appropriate place
in the surface Brillouin zone \cite{Pai94}.  However, the evidence for the influence
of surface states was not compelling due to the large number of fitting parameters
compared to the amount of experimental data.  
 
Convincing evidence of long-range, surface-state
mediated interactions between Cu atoms on Cu\{111\} has just appeared \cite{RR00}.  
It is tempting to invoke these interactions (which could even decay as $\ell^{-3/2}$ due to
the isotropy of the state \cite{RZ92}) as the source of the large-$s$ humps in ``poor" data on
this surface.  However, this idea does not explain similar ``poor" data on Cu\{100\}, where
the image states are far from the Fermi energy.

The energy of the long-range interaction measured for Cu atoms on Cu\{111\} is notably 
weak \cite{RR00}: the deepest minimum corresponds to an attraction of 0.4 meV at 27\AA.  However,
the consequent prefactor of the expression in Eq.~(\ref{e:indir}) is 0.3 eV-\AA$^2$, which is
comparable to the values of $A$ listed in Table 2 (if some length of order an atomic spacing is
used to adjust the units).

The transport properties of fractional metal overlayers have received close scrutiny in
recent years \cite{HTTSN}.  Since the metallic surface states can be tuned in these systems,
it is intriguing to speculate about engineering morphology using such step interactions.  

\section{Concluding Remarks}
\label{sec:conclu}

The CGWD of
Eq.~(\ref{e:Ps}) is  an excellent interpolation between the established points
at $\tilde{A}$=0 and
$\tilde{A}$=2, and approaches the correct limit for very
large $\tilde{A}$.  Qualitatively it certainly captures the global
behavior of variance as a function of $\tilde{A}$, and numerical evidence suggests that it
interpolates well in the regime of large $\tilde{A}$.  While the shape of the TWD does
approach a Gaussian in this regime of moderately strong
$\tilde{A}$, the CGWD (via Eq.~(\ref{e:as}) provides arguably the best way to extract $\tilde{A}$
from the variance of the TWD.  Of the several ways to extract $\tilde{A}$ from fits to a
Gaussian, the Saclay (R) scheme is better for moderate $\tilde{A}$ while the Grenoble (EA(all))
scheme is better for stronger $\tilde{A}$.  

The difficulty of accurately estimating the value $\tilde{A}$ from the TWD, especially from
its variance or width, is exacerbated by the extreme sensitivity of $\sigma^2$ to $\tilde{A}$:
fractional errors in the deduced widths of TWDs are magnified by a factor of 4 in
$\tilde{A}$.  For many applications, determining the relative size of the step-step repulsions
between different systems is more important than deducing their absolute sizes; in such
comparisons, it is crucial that the analysis of
$\tilde{A}$ be done using the same approach for all systems.  (Likewise, experimentalists
should state clearly the raw [dimensionless] width,
$\sigma$---or the value of $\varrho$ in a fit to the CGWD.)  

Often the extracted value is
rationalized by misapplying the celebrated result of Marchenko and
Parshin\cite{mp} relating the step repulsions to surface stress \cite{mp}. That formula
assumes an elastically isotropic substrate and asymptotically large separations.
Usually one, often both, of these conditions do not apply, and there is no well-prescribed
procedure to compute corrections.  Furthermore, the in-plane component of the stress
dipole is not measurable and is often neglected. Thus, establishing quantitative
connections between deduced $\tilde{A}$ and surface stresses is [even] harder than
extracting reliable quantitative estimates of $\tilde{A}$. 

Another worrisome assumption is that the step interactions are ``instantaneous"
in the 1+1 D perspective (i.e.\ occur only between points on steps at the same
coordinate along the mean step direction $\hat{y}$) becomes particularly
questionable when steps are close together and have large wandering
fluctuations with short wavelengths.  In the latter situation, the description
of single-step fluctuations in terms of stiffness may also break down.

In systems in which surface states near the Fermi energy play an active role, there should be
notable effects on the TWD and the consequent surface morphology.  Multistep correlations have
received little attention, even though data is readily available in experiments measuring TWDs. 
Moreover, it is almost as easy to tabulate the step-step pair correlation function as the TWD,
but easier to decipher theoretically.  We have provided several hints and warnings, hopefully
useful, for experimentalists studying spacings on vicinals.

Most theoretical activity dealing with random fluctuations in complex systems has focused on the
the three special values of $\varrho$, and occasionally on interpolations between them.  This
corresponds to weak $\tilde{A}$.  The few exceptions focus on pair
correlations and are rather technical
\cite{f93}.  More explicit numerical investigations in this regime would be illuminating. 
Moreover, there remains the mystery of why the CGWD works so well when there is no fundamental
symmetry argument to justify it.

The Calogero-Sutherland model has been termed an ideal \cite{SB99} Luttinger liquid 
\cite{HCP98,V00}. 
Connections have been made to edge states in the quantum Hall effect \cite{K93}, non-linear waves
in a stratified fluid
\cite{CLP}, and a host of more abstract problems.  Most of these systems exhibit
corrections, making it difficult to make detailed connections with the theory \cite{SB99}.  It
will be interesting to see whether similar problems involving corrections to the $A/\ell^2$ much
discussed above confound a similar effort for vicinal surfaces.  Furthermore, many of the
interesting properties involve dynamic correlations, which for vicinal surfaces translates to
correlations between displacements on [different] steps at {\it different} values of $y$. In any
case, however, these connections between the properties of vicinal surfaces and other
active fields add to the fascination of the subject.

\vspace{-0.5cm}

\section*{Acknowledgment}

Work was supported by the NSF-MRSEC at University of Maryland,
done in collaboration with N.~C. Bartelt, B. Jo\'os, E.~D. Williams, J.~E.
Reutt-Robey, et al.\ at UM, and with M. Giesen and H. Ibach at FZ-J\"ulich (via a
Humboldt U.S. Senior Scientist Award).  SDC participated in an NSF-MRSEC-sponsored
REU program. 

\vspace{-.7cm}


\onecolumn

\noindent Table 1:
 Tabulation of predictions of the variance of terrace-width
distributions $P(s)$ [where $s$ is the terrace width normalized by its
average value] based on exact results at the three soluble values of the
dimensionless interaction strength,
$\tilde{A}$, the corresponding Wigner-surmise expression, and several ways of
interpreting a Gaussian fit.
\vspace{0.2cm}

\begin{center}
\begin{tabular}{ccc|c|cccc} \hline 
\hline Property & Case & Abbrev.\ & Ref.\ &
$\varrho =2$ &
$\varrho =4$ & Arbitrary $\varrho$ &$\varrho \rightarrow \infty$\\
    & & X & & Non-interact & Exact rpl. & Repulsive & Extreme rpl. \\
\multicolumn{2}{c}{$\varrho = 1 + \sqrt{1 +4\tilde{A}}
\equiv 2 \surd\tilde{A}_{\rm eff}$} 
& & & $\tilde{A} =0$ &  $\tilde{A}=2$&$\tilde{A} \! = \! (\varrho \! - \! 2)\varrho/4$ & 
$\tilde{A} \! \rightarrow \! \tilde{A}_{\rm eff} \! \equiv \! \varrho^2/4$ \\  \hline
 \multicolumn{2}{c} {Symmetry assoc. w/ Sutherland $\cal H$} & &
& unitary  & symplectic & &  [SHO+phonons] \\
 \hline
 \multicolumn{2}{l}{$a_{\varrho}= 2\left[\Gamma
\left(\frac{\varrho +2}{2}\right)\right]^{\varrho +1}/
                   \left[\Gamma
\left(\frac{\varrho +1}{2}\right)\right]^{\varrho +2} $} & &
\onlinecite{MehtaRanMat,Guhr98} &
$32/\pi^2$ &
$(64/9 \pi)^3$& In leftmost & $2\exp[(\varrho+3)/2]$\\ 
 \multicolumn{2}{l}{$b_{\varrho} = 
  \left[\Gamma \left(\frac{\varrho +2}{2}\right)/
             \Gamma \left(\frac{\varrho + 1}{2}\right)\right]^2$} & & & $4/\pi$
 & $64/9 \pi$& column & $\varrho/2+1/4$\\
\hline Variance& Exact  & [all] & \onlinecite{MehtaRanMat,Guhr98} & 0.180 & 0.105 & ---
& $0.495/\varrho$ \\
$\sigma^2 =\mu_2$&Wigner surmise & W[all] & \cite{EP99,MehtaRanMat,Haake91} &0.1781 & 0.1045 &
$(\varrho\! +\! 1)/2b_{\varrho} -\! 1$&$0.500/\varrho$\\
\cline{2-8}
 \hspace{0.2cm} $=\mu^{\prime}_2 -1$& {\it Gruber-Mullins} & GM(all) &
\onlinecite{Gruber67} & 0.1307 & 0.0981 & $0.139 /\surd\tilde{A}$&$0.278/\varrho$\\ 
 & {\tt "} & GM(NN)  & {\tt "} & 0.1307 & 0.1021 & $0.144 /\surd\tilde{A}$
&$0.289/\varrho$\\ 
{\it Gaussian} & {\it modified Grenoble}  &EA(all) & \onlinecite{EP99,PM98,IMP98} & 0.247 &
0.1185 & 
$0.247 /\surd\tilde{A}_{\rm eff}$&$0.495/\varrho$\\
 {\it alternatives} & {\tt "} & EA(NN) & {\tt "} & 0.260 & 0.1300 & 
$0.260 /\surd\tilde{A}_{\rm eff}$&$0.520/\varrho$\\ 
  & {\it Saclay} & R  &
\onlinecite{Masson94,Barbier96,LeGoff99} & 0.203 & 0.101 &
$ 0.203 /\surd\tilde{A}_{\rm eff}$&$0.405/\varrho$\\
\hline
Neighboring&Exact  $\langle (s_1 + s_2 -2)^2\rangle$ & &\onlinecite{MehtaRanMat}  &
 0.248 & 0.138 & &$0^+$\\
 terraces& Multistep CGWD  & & \cite{AS99} & 0.257 &
0.145 & &$0^+$\\ \hline
\hline  
\end{tabular}
\end{center}

\vspace{0.1cm}

The ``unmodified" Grenoble expression \cite{PM98,IMP98} is obtained by substituting $\tilde{A}$
for $\tilde{A}_{\rm eff}$. The entries for the variance at $\tilde{A}$=2 are almost
50\% larger.  The bracketed ``[all]" is a reminder that the Calogero-Sutherland model---and, 
hence, the Exact solutions and the Wigner surmises---involves all steps interacting.  As
$\tilde{A}$ increases, the TWD becomes narrower, more symmetric, and more nearly
Gaussian.  Anticorrelations of neighboring terrace-width fluctuations increase.  For
the three exactly-solvable [non-trivial] cases, the Wigner surmise provides an
excellent approximation, significantly better than any alternative.  

The covariance of the fluctuations of neighboring steps is the
difference from unity of the tabulated correlation $\langle (s_1 + s_2 -2)^2\rangle$
divided by twice the variance.  It is negative, indicating that fluctuations of
adjacent terrace widths are anticorrelated.
 For Gruber-Mullins this
covariance is {\it ipso facto} $-1$.
With the Grenoble formalism\cite{PM98,IMP98} we find it to be --1/3 (NN) or
--0.36$\ldots$ (all); with the Saclay formalism \cite{Barbier96},
it is  --0.33.  In all these cases, the covariance is
independent of $\tilde{A}$. In contrast, the exact covariance increases weakly in
magnitude with $\tilde{A}$, from --0.31 at
$\tilde{A}$= 0 to --0.34 at $\tilde{A}$= 2 \cite{MehtaRanMat}, and presumably to
--0.36 asymptotically.

\newpage

\noindent Table 2: Compendium of experiments measuring the variances of terrace width
distributions of vicinal systems. 
\begin{center}
\begin{tabular}{cc|c|cc|c|c|c} \hline 
Vicinal & T(K) & $\sigma^2$ & $\varrho$ & $\tilde{A}$ &$A_W/A_G$&
$A_W$ (eV-\AA)& Experimenters \\
\hline
Pt(110)-(1$\times$2) & 298 & & 2.2 & 0.13 &---&$\tilde{\beta}=?$&Swamy, Bertel
\cite{SBV} \\ Cu (19,17,17) & 353 & 0.122 & 4.1 & 2.2 & 0.77& 0.005& Giesen
\cite{RCEG00,GS99}
\\ Si(111) & 1173 & 0.11 & 3.8 &
1.7 & 0.96& 0.4  &Bermond, M\'etois \cite{BMHF} \\
Cu(1,1,13) & 348 & 0.091 & 4.8 & 3.0 &1.27&0.007 &Giesen \cite{RCEG00,G97} \\
Cu(11,7,7) & 306 & 0.085 & 5.1 & 4 &1.37& 0.004 &Giesen \cite{RCEG00,GS99} \\
Cu(111) & 313 & 0.084 & 5.0 & 3.6 &1.39&0.004 &Giesen \cite{RCEG00,GS99} \\
Cu(111) & 301 & 0.073 & 6.0 & 6.0 &1.58&0.006& Giesen \cite{RCEG00,GS99} \\
Ag(100) & 300 & 0.073 & 6.4 & 6.9 &1.58&$\tilde{\beta}=?$ &P. Wang$\ldots$Williams \\
Cu(1,1,19) & 320 & 0.070 & 6.7 & 7.9 &1.64&0.012 &Giesen \cite{RCEG00,G97} \\
Si(111)-(7$\times$7) & 1100 & 0.068 & 6.4 & 7.0 & 1.67& 0.7&Williams \cite{W94} \\ 
Si(111)-(1$\times$1)Br & 853 & 0.068 & 6.4 & 7.0 & 1.67&0.1 &X.-S. Wang,
Williams \cite{WW}\\ 
Si(111)-Ga  & 823 & 0.068 & 6.6 & 7.6 &1.67& 1.8 & Fujita...Ichikawa
\cite{FKI}\\ 
Si(111)-Al $\surd3$ & 1040 & 0.058 & 7.6 & 10.5 & 1.85&2.2& 
Schwennicke$\ldots$Williams \cite{SW}\\ 
Cu(1,1,11) & 300 & 0.053 & 8.7 & 15 & 1.95&
0.02& Barbier et al.\ \cite{Barbier96}
\\ Cu(1,1,13) & 285 & 0.044 & 10 & 20 & 2.12&0.02 &Giesen \cite{RCEG00,G97}\\  
Pt(111) & 900 & 0.020 & 24 & 135 &2.59&6& Hahn$\ldots$Kern \cite{HSMFK}\\ 
Si(113) rotated & 1200 & 0.004 & 124 & 3.8$\times$10$^3$ & 2.92&
(27$\pm$5)$\times$10$^2$& van Dijken,Zandvliet,Poelsema \cite{DZP} \\
\hline  
\end{tabular}
\end{center}

\vspace{0.1cm}

The estimate of $\tilde{A}$
is obtained from the (normalized) variance using Eq.~(\ref{e:as}), except for the first-row entry,
which is based on a direct fit using the 2-parameter CGWD. $A_G$ is short for
$A_{GM(NN)}$, the conventional estimate (cf.\ Table 7 of Ref. \cite{JW99}).  The
column ``$A_W/A_G$", computed using Eq.~(\ref{e:aa}), shows that for most systems
the correction factor is of order half the ultimate asymptotic factor.  For the
copper entries, $A_W$ is computed from Eq.~(\ref{e:defAtilde}), using
$\tilde{\beta}_{TSK} = 2k_BT(a_{\parallel}/a^2_{\perp})\sinh^2(\epsilon/2k_BT)$ and
kink energies $\epsilon$ of 0.126 and 0.12 for vicinals to \{100\} [nominally
(1,1,2$n+1$)] and \{111\}, respectively.  In other cases, $A_W$ is simply rescaled
from Ref.~\cite{JW99}.  In two cases, the values of the stiffness are not readily
available.

\end{document}